\documentclass[11pt]{article}

\usepackage{amsmath}    
\usepackage{amssymb,amsbsy}     
\usepackage[dvips]{epsfig}      
\usepackage{pstricks}

\makeatletter
\def\@listI{\leftmargin\leftmargini \parsep 0pt\topsep
3pt plus 1pt minus 1pt\itemsep 2pt}
\makeatother

\input{macros2e97}
\renewcommand\P{{\rm P}}

\newcommand\NL{{\rm NL}}

\newcommand\SACe{{\rm SAC}^1}
\newcommand\NCe{{\rm NC}^1}
\newcommand\ACn{{\rm AC}^0}
\newcommand\ACCn{{\rm ACC}^0}
\newcommand\TCn{{\rm TC}^0}

\newcommand\LOGCFL{{\rm LOGCFL}}

\newcommand\N{{\mathbb{N}}}	

\newcommand\set[2]{\bigl\{\,#1\bigm| #2\,\bigr\}}
\newcommand\seq{\subseteq}

\newcommand\eqdef{=_{\rm def}}

\newcommand\quark{\kern.5em}

\newcommand\CFL{{\rm CFL}}
\newcommand\REG{{\rm REG}}

\newcommand {\nat}      {\mathbb{N}}

\newcommand\FO{{\rm FO}}

\newcommand\SOM{{\rm MSO}}

\newcommand\posFO{{\rm FO}^+}

\newcommand\calA{{\mathcal{A}}}

\newcommand\Grp{\text{\rm Grp}}
\newcommand\QB{Q_B}
\newcommand\QG{Q_G}
\newcommand\QGrp{Q_{\Grp}}
\newcommand\bit{_{\text{\rm bit}}}
\newcommand\FObit{\FO\bit}

\newcommand\un{^{\text{\rm un}}}
\newcommand{\Qun}{\QGrp\un}

\newcommand\BC[1]{{\rm BC}(#1)}
\newcommand\posBC[1]{{\rm BC}^+(#1)}

\newcommand{\PBS}[1]{\let\temp=\\#1\let\\=\temp}

\newcommand{\ol}[1]{\overline{#1}}

\begin{document}


\title{\Large\bf The Descriptive Complexity Approach to LOGCFL}

\author{Clemens Lautemann%
\thanks{Institut f\"ur Informatik, Johannes-Gutenberg-Universit\"at Mainz,
55099 Mainz, Germany.},
Pierre McKenzie%
\thanks{Informatique et recherche op{\'e}rationnelle,
Universit{\'e} de Montr{\'e}al, C.P.\ 6128, Succ.\ Centre-Ville,
Montr{\'e}al (Qu{\'e}bec), H3C 3J7 Canada.
Research performed while on leave at the Universit\"at T\"ubingen.
Supported by the (German) DFG, the (Canadian) NSERC and the (Qu\'ebec)
FCAR.},\\[1ex]
Thomas Schwentick\footnotemark[1],
Heribert Vollmer%
\thanks{Theoretische Informatik, Universit\"at W\"urzburg,
Am Exerzierplatz 3, 97072 W\"urzburg, Germany.}}

\date{}
\maketitle

%
\begin{abstract}
Building upon the known generalized-quantifier-based first-order
characterization of \LOGCFL, we lay the groundwork for a deeper
investigation.
Specifically, we examine subclasses of \LOGCFL\ arising from varying
the arity and nesting of {\em groupoidal} quantifiers.
Our work extends the elaborate theory relating {\em monoidal}
quantifiers to $\NCe$ and its subclasses.
In the absence of the BIT predicate, we resolve the main issues:
we show in particular that no single outermost unary groupoidal
quantifier with FO can capture all the context-free languages, and we obtain
the surprising result that
a variant of Greibach's ``hardest context-free
language'' is \LOGCFL-complete under quantifier-free BIT-free projections.
We then prove that FO with unary groupoidal quantifiers
is strictly more expressive with the BIT predicate than without.
Considering a particular groupoidal quantifier, we prove that first-order
logic with majority of pairs is strictly more expressive than first-order with
majority of individuals.
As a technical tool of independent interest,
we define the notion of an aperiodic nondeterministic finite
automaton and prove that FO translations are precisely the
mappings computed by single-valued aperiodic
nondeterministic finite transducers.\\[-2mm]

\end{abstract}
%

\section{Introduction}

In {\em Finite Automata, Formal Logic, and Circuit
Complexity} \cite{str94}, Howard Straubing surveys an
elegant theory relating
finite semigroup theory, first-order logic, and computational complexity.
The gist of this theory
is that questions about the structure of the complexity class
$\NCe$, defined from logarithmic depth bounded fan-in Boolean
circuits,
can be translated back and forth into questions about the
expressibility of first-order logic augmented with
new predicates and quantifiers.
Such a translation provides new insights, makes tools from one field
available in the other, suggests tractable refinements to the hard
open questions in the separate fields, and puts the
obstacles to further progress in a clear perspective.

In this way, although, for example,
the unresolved strict containment in $\NCe$ of
the class $\ACCn$, defined from bounded-depth polynomial-size unbounded
fan-in circuits over \{AND, OR, MOD\}, remains a barrier
since the work of Smolensky \cite{smo87},
significant progress was made in (1) understanding the power of the BIT
predicate and
the related circuit uniformity issues \cite{baimst90}, (2) describing
the regular languages within subclasses of $\NCe$
\cite{bacostth92,mcpeth91}, and (3) identifying 
the all-important role of the interplay between arbitrary and
regular numerical predicates in the status of the $\ACCn$ versus
$\NCe$ question \cite[p.~169, Conjecture IX.3.4]{str94}.

Barrington, Immerman and
Straubing \cite{baimst90} introduced the notion of a
{\em monoidal} quantifier and noted that, for any non-solvable
group $G$, the class $\NCe$ can be described using first-order logic
augmented with a monoidal quantifier for $G$.
Loosely speaking,
such a quantifier 
provides a constrained ``oracle call'' to the word problem for $G$
(defined essentially as the problem of computing the product of a
sequence of elements of $G$).

B\'edard, Lemieux and McKenzie \cite{belemc93} later noted that there
is a fixed finite {\em groupoid}
whose word problem
is complete for the class \LOGCFL\ of
languages reducible in logarithmic space to a context-free language
\cite{coo71,sud78}.
A groupoid $G$ is a set with a binary operation
satisfying no discernible property, and the word problem for $G$
is that of computing the set of all legally bracketed products of
a given sequence of elements of $G$.
It is not hard to see that any context-free language is the word
problem of some groupoid, and that any groupoid word problem is
context-free (see \cite[Lemma 3.1]{belemc93}).

It followed that \LOGCFL,
a well-studied class which contains
nondeterministic logarithmic space \cite{sud78}
and is presumably much larger than
$\NCe$, can be described by first-order logic
augmented with {\em groupoidal} quantifiers.
These quantifiers can be defined formally as Lindstr\"om quantifiers
\cite{lin66} for context-free languages.

In this paper, we take up the groupoidal first-order characterization
of \LOGCFL, and initiate an investigation of \LOGCFL\ from the
viewpoint of descriptive complexity.  The rationale for this study,
which encompasses the study of $\NCe$, is that tools from logic might
be of use in ultimately elucidating the structure of \LOGCFL.  We do
not claim new separations of the major subclasses of \LOGCFL\ here.
But we make a first step, in effect settling necessary preliminary
questions afforded by the first-order framework.

Our precise results concern the relative expressiveness of first-order
formulas with ordering (written FO), interpreted over finite strings,
and with: (1) 
nested versus unnested groupoidal quantifiers, 
(2) unary versus non-unary groupoidal quantifiers,
(3) the presence versus the absence of the BIT predicate.
Feature (3) was the focus of an important part of the work by
Barrington, Immerman and Straubing \cite{baimst90} on uniformity
within $\NCe$.
Feature (2) was also considered, to a lesser extent, by the same
authors, who left open the question of whether the
``majority-of-pairs'' quantifier could be simulated by a
unary majority quantifier  in the absence of the BIT predicate
\cite[p.~297]{baimst90}.
Feature (1) is akin to comparing many-one
reducibility with Turing reducibility in traditional complexity theory.

Here we examine all combinations of features (1), (2) and
(3).
Our separation results are summarized on Fig.~\ref{incdiag}
on p.~\pageref{incdiag}.
In the absence of the BIT predicate, we are able to
determine the following relationships:
\begin{itemize}
\item $\FO$ to which a single unary groupoidal quantifier is applied,
written 
$\QGrp\un\FO$, captures the CFLs,
and is strictly less expressive than FO with nested unary quantifiers,
written $\FO(\QGrp\un)$, which in its turn is strictly weaker than \LOGCFL.
A consequence of this result, as we will see, is an answer to
the above mentioned open question from \cite{baimst90}:
We show that first-order with the majority-of-pairs quantifier
is strictly more expressive than
first-order logic with majority of individuals.
\item No single groupoid $G$ captures all the CFLs
as $\QG\un\FO$, i.\,e.~as FO to which the single unary groupoidal
quantifier $\QG\un$ is applied, 
\item $\FO$ to which a single {\em non-unary} groupoidal quantifier is
applied, 
written $\QGrp\FO$, captures \LOGCFL;
our proof implies,
remarkably, that adding a padding symbol to Greibach's hardest
context-free language
\cite{gre73}, see also \cite{aubebo97},
yields a language which is \LOGCFL-complete under BIT-free quantifier-free
projections.
\end{itemize}
When the BIT predicate is present,
first-order with non-unary
groupoidal quantifiers of course still describes \LOGCFL.
In the setting of monoidal quantifiers \cite{baimst90}, FO with BIT is
known to capture uniform circuit classes, notably uniform $\ACCn$,
which have not yet been separated from $\NCe$.
We face a similar situation here: the BIT predicate allows capturing
classes (for example $\FO\bit(\QGrp\un)$, verifying
$\TCn\subseteq\FO\bit(\QGrp\un)\subseteq\LOGCFL$),
which only a major breakthrough would seem to allow
separating from each other.
We are able to attest to the strength of the BIT predicate
in the setting of unary quantifiers, proving that:
\begin{itemize}
\item $\QGrp\un\FO\subsetneq\QGrp\un\FO\bit$, i.\,e.~(trivially) some
non-context-free
languages are expressible using BIT and a single unary groupoidal
quantifier,
\item $\FO(\QGrp\un)\subsetneq\FO\bit(\QGrp\un)$, i.\,e.~(more
interestingly) BIT adds
expressivity even when unary groupoidal quantifiers can be nested.
\end{itemize}

We also develop a technical tool of independent interest, in the form
of an aperiodic (a.\,k.\,a.~group-free, a.\,k.\,a.~counter-free)
nondeterministic finite automaton.  Aperiodicity has
been studied intensively, most notably in connection with the star-free
regular languages \cite{sch65}, but, to the best of our knowledge,
always in a deterministic
context.  Here we define a NFA $A$ to be
aperiodic if the DFA resulting from
applying the subset construction to $A$ is aperiodic.  The usefulness
of this notion lies in the fact, proved here, that first-order
translations are precisely those mappings which are computable by
single-valued aperiodic nondeterministic finite transducers.

Section 2 in this paper describes our first-order framework and
exhibits a link between
standard formal language operations and
unary generalized quantifiers.
Section 3 introduces nondeterministic finite transducers and proves
that they characterize first-order translations.
Section 4 forms the bulk of the paper and
develops the relationships between our logic-based \LOGCFL\ subclasses.
Section 5 concludes
with a number of suggestions how to extend the results obtained here.



\section{Preliminaries} \label{sectprelim}

\subsection{Complexity theory}

$\REG$ and $\CFL$ refer to the regular and to the $\epsilon$-free
context-free languages respectively.
The CFL results in this paper could be adapted to treat the empty
string $\epsilon$ in standard ways.
We will make scant reference to the inclusion chain
$$\ACn \subsetneq \ACCn \subseteq \TCn \subseteq \NCe \subseteq \NL
\subseteq \LOGCFL = \SACe \subseteq \P,$$
where we assume familiarity with $\NCe$, \NL, and \P, and recall that
\begin{itemize}
\item $\ACn$ (resp.~$\ACCn$) (resp.~$\TCn$) is the set of
languages
recognized by sufficiently uniform families of constant depth,
polynomial size, unbounded fan-in circuits over the basis
$\{ \wedge, \vee, 
\neg \}$ (resp.~over a basis consisting of $\{ \wedge, \vee \}$ together
with a single Boolean MOD$_q$ gate, defined to output $0$ iff $q$
divides the sum of its input bits)
(resp.~over the basis consisting solely of $\neg$ and the
{\rm MAJORITY} gate, defined to output $1$ iff at least half of
its input bits are set),

\item \LOGCFL\ is the set of languages logspace-reducible to a
context-free language \cite{coo71,sud78}; alternatively, this class is
$\SACe$, namely the set of languages recognized by uniform 
families of log depth, polynomial size,
Boolean circuits in which $\wedge$ has bounded fan-in and the fan-in
of $\vee$ is unrestricted \cite{ven91}.

\end{itemize}

\subsection{The first-order framework}

We consider first-order logic with linear order.
We restrict our attention to 
{\em string signatures}, i.\,e.~signatures of the form
$\langle P_{a_1},\dots,P_{a_s}\rangle$, where all the predicates
$P_{a_i}$ are unary, and in every 
structure $\calA$, $\calA\models P_{a_i}(j)$ iff the $j$th symbol in the input 
is the letter $a_i$.
Such structures are thus
words over the alphabet $(a_1,\dots,a_s)$,
and first-order variables range over positions within such
a word,
i.\,e.~from $1$ to the word length $n$. For technical reasons that 
will become apparent shortly, we assume here,
as in the rest of the paper, a linear order on each alphabet and we 
write alphabets as sequences of symbols to indicate that order.

Our basic formulas are built from variables in the usual way,
using the Boolean connectives $\{ \wedge, \vee, \neg \}$, the relevant
predicates $P_{a_i}$ together with $\{ = , <\}$,
the constants min and max,
the quantifiers $\{\exists, \forall \}$, and parentheses.
We will occasionally use the binary predicate
${\rm BIT}(x,y)$, defined to be true iff the $x$th bit in the binary
representation of $y$ is $1$.
We write $\BC{\cal L}$ to denote the Boolean closure of the set $\cal L$ of
languages 
(i.\,e.~closure under intersection, union, and complement)
and $\posBC{\cal L}$ to denote the closure under union and
intersection only.

\begin{defi} \label{grquant}
{\em Lindstr\"om quantifier.}
Consider a language $L$ over an alphabet $\Sigma = (a_1, a_2,
\ldots, a_{s})$.
Let $\ol{x}$ be a $k$-tuple of variables (each of which
ranges from $1$ to the ``input length'' $n$, as we have seen).
In the following, we assume the
lexical ordering on $\{1,2,\ldots,n\}^k$,
and we write $X_1, X_2, \ldots , X_{n^k}$ for the sequence of
potential values taken on by $\ol{x}$.
The groupoidal quantifier $Q_L$ binding $\ol{x}$ takes a
meaning if $s-1$ formulas, each having as free variables
the variables in $\ol{x}$ (and possibly others), are available.
Let $\phi_1(\ol{x})$, 
$\phi_2(\ol{x})$, $\ldots$,
$\phi_{s-1}(\ol{x})$ be these $s-1$ formulas.
Then $Q_L\ol{x}\bigl[
\phi_1(\ol{x}),
\phi_2(\ol{x}), \ldots,
\phi_{s-1}(\ol{x})\bigr]$ holds on a string $w=w_1\cdots w_n$, iff the word of length $n^k$ whose $i$th
letter, $1\leq i\leq n^k$, is
\begin{displaymath}
\left\{
   \begin{array}{ll}
       a_1 & \mbox{if $w \models \phi_1(X_i)$,} \\
       a_2 & \mbox{if $w \models \neg\phi_1(X_i)\wedge\phi_2(X_i)$,} \\
        &\dots\\
       a_{s}& \mbox{if  $w \models \neg\phi_1(X_i) \wedge\neg
   \phi_2(X_i) \wedge\ldots\wedge \neg\phi_{s-1}(X_i)$,}   
   \\ 
   \end{array}
       \right.
\end{displaymath}
belongs to $L$.
Thus the formulas $[\phi_1(\ol{x}),\phi_2(\ol{x}), \ldots,
\phi_{s-1}(\ol{x})]$ fix a function mapping an input word/structure $w$
of length $n$ to a word of length $n^k$. This function is called the
{\em reduction\/} or {\em transformation\/} defined by 
$[\phi_1(\ol{x}),\phi_2(\ol{x}), \ldots,\phi_{s-1}(\ol{x})]$.
In case we deal with the binary alphabet ($s=2$) we omit the
braces and write $Q_L\ol{x}\phi(\ol{x})$ for short.
\end{defi}

\begin{defi}
A {\em groupoidal quantifier\/} is a Lindstr\"om quantifier
$Q_L$ where $L$ is a context-free language.
\end{defi}

The Lindstr\"om quantifiers of Definition \ref{grquant} are
more precisely what has been refered to as 
``Lindstr\"om quantifiers on string'' \cite{buvo96}.
The original more general definition \cite{lin66} uses transformations to
arbitrary structures, not necessarily of string signature.
However, in the context of this paper reductions to CFLs play a role of
utmost importance, and hence the above definition seems to be the most
natural.

The terminology ``groupoidal quantifier'' stems from the fact
that any context-free language is a
word problem over some groupoid \cite[Lemma~3.1]{belemc93},
and vice-versa every word problem of a finite groupoid is context-free.
Thus a Lindstr\"om quantifier on strings defined by a context-free
language is nothing else than a Lindstr\"om quantifier (in the classical
sense) defined by a structure that is a finite groupoid multiplication table.

Barrington, Immerman, and Straubing, defining {\em monoidal quantifiers\/}
in \cite{baimst90}, in fact proceed along the same avenue: they
first show how monoid word problems can be seen as 
languages, and then define generalized quantifiers given by such languages
(see \cite[pp.~284f.]{baimst90}).

We refer the reader to standard texts for formal details on the
semantics of our logical framework.  For instance, Definition
\ref{grquant} skims over the semantics of a groupoidal quantifier in
the case in which the underlying formulas contain free variables other
than those in $\ol{x}$.  We find Straubing's handling of these issues
\cite{str94} particularly convenient and we will occasionally refer to his
treatment.

\subsection{Groupoid-based language classes}

Here we define our first-order language classes precisely.
Fix a finite groupoid $G$.
Each $S\subseteq G$ defines a language ${\cal W}(S,G)$ composed of all
words $w$, 
over the alphabet $G$, which ``multiply out'' to an element of $S$
when an appropriate legal bracketing of $w$ is chosen.

\begin{defi} 
$\QG\FO$ is the set of languages describable by applying
a single groupoidal quantifier $Q_L$ to an appropriate tuple of FO
formulas, where $L={\cal W}(S,G)$ for some $S\subseteq G$.\\
$\QGrp\FO$ is the union, over each finite groupoid $G$, of $\QG\FO$.\\
$\FO(\QG)$ and $\FO(\QGrp)$ are defined
analogously, but allowing groupoidal quantifiers to be used as any
other quantifier would (i.\,e.~allowing arbitrary nesting).\\
$\QG\un\FO$ and $\FO\bit(\QGrp\un)$, etc, are defined analogously,
but possibly allowing the BIT predicate (signaled by subscripting FO
with bit) and/or restricting to
{\em unary} groupoidal quantifiers (signaled by the exponent
``un'').
\end{defi}

We use $\FO(+)$ to denote that the additional predicate
``$x+y=z$'' (with the obvious semantics) is additionally allowed.
It is known that $\FO(+)$ can express exactly the semi-linear
sets (see \cite[p.~231]{har78}).

\subsection{Unary quantifiers and homomorphisms}

We will encounter unary groupoidal quantifiers repeatedly.
Here we show how these relate to
standard formal language operations.  Recall that a length-preserving
homomorphism $\Sigma^*\rightarrow\Delta^*$ is the unique free monoid
morphism extending a map $h\colon\Sigma\rightarrow \Delta$ for
finite alphabets $\Sigma,\Delta$.
In a different context, a result very similar to the next theorem
is known as {\em Nivat's Theorem\/}
\cite[Theorem 3.8, p.~207]{masa97}.


\begin{thm}\label{Nivat}
Let $B$ be an arbitrary language, and let $A$ be describable in
$\QB\un\FO$,
that is, by a first order sentence preceded by one unary Lindstr\"om
quantifier 
(i.\,e.~binding exactly one variable).
Then there are length-preserving homomorphisms $g,h$
and a regular language $D$ such that $A=h(D\cap g^{-1}(B))$.
\end{thm}

\begin{proof}
\newcommand\bigletter[2]{\genfrac[]{0pt}{}{#1}{#2}}
Let $A$ be defined by the formula
$\psi\in\QB\un\FO$, $\psi=\QB x\phi(x)$, $B\seq\Gamma^*$
(assuming $\Gamma=(0,1)$ initially).
Let $\Delta$ be the underlying alphabet determined by the string signature.
$\phi$ thus defines a mapping from words over $\Delta$ to binary words.
Define $D$ to consist of all words
$\bigletter{u_1}{y_1}\cdots\bigletter{u_k}{y_k}$ such that
$\phi$ maps $u_1\cdots u_k$ to $y_1\cdots y_k$.
Define the homomorphisms $h$ and $g$ 
by $h\colon\bigletter{a}{b}\mapsto a$ 
and $g\colon\bigletter{a}{b}\mapsto b$
for all $a\in \Delta$ and  $b\in\Gamma$.
Then $h(D\cap g^{-1}(B)) = A$.
But why is $D$ regular?  Intuitively, $D$ is regular because FO languages
are regular.  Arguing formally requires a bit of
care because each 
$y_i$ depends on the truth value of an FO formula in which the
variable $x$ is instantiated with $i$.  A proof that a finite
automaton is able to determine $y_i$ can be found in Straubing
\cite[pp.~23--24]{str94}.  To see that $D$ itself is regular, note that
an NFA $N$ can guess an incorrect $y_i$ (by guessing the position of
the formal variable $x$ in a $\cal V$-structure, borrowing
notation from Straubing) and verify that $y_i$ is incorrect.
In this way $N$ accepts the complement of $D$, so that $D$ is
regular\footnote{An alternative proof that $D$ is regular is immediate
from Theorem \ref{FO<->NFA}.}.

The above strategy to show the regularity of $D$ adapts to the
case of a non-binary alphabet $\Gamma$, in which case $N$ is a
direct product of the NFAs accepting the languages defined by the
relevant tuple of FO formulas.
The homomorphisms $g$ and $h$ are unchanged.
\end{proof}

\begin{rem}\rm
$\FO$ precisely captures the variety of star-free regular languages
\cite{mcpa71}, which allows us to even conclude that the $D$ above
is star-free.
\end{rem}

\section{An automaton characterization of FO-translations}

As a technical tool, it will be convenient to have an
automata-theoretic characterization of {\em first-order translations}, 
i.\,e.~of reductions defined by $\FO$-formulas with one free variable.
Since $\FO$ precisely describes the (regular) languages accepted by
aperiodic deterministic finite automata \cite{mcpa71}, one might
expect aperiodic deterministic finite transducers to capture
$\FO$-translations.
This is not the case however because, e.g.\ the FO-translation which
maps every string $w_1\cdots w_n$ to $w_n^n$ cannot be computed by
such a device.

We show in this section that the appropriate automaton model to use is
that of a single-valued aperiodic nondeterministic finite transducer,
which we define and associate with $\FO$-translations in this section.
But first, we discuss the notion of an aperiodic NFA.

\begin{defi} \label{defaper}
A deterministic or nondeterministic FA $M$ is {\em aperiodic}
(or group-free) iff there is an $n\in\N$ such that for all 
states $s$ and all words $w$,
$$\delta(s,w^n) = \delta(s,w^{n+1}).$$
Here $\delta$ is the extension of $M$'s transition function from
symbols to words. Observe that if $M$ is nondeterministic then
$\delta(t,v)$ is a set of states, i.\,e.~locally here we abuse notation by
not distinguishing between $M$'s extended transition function $\delta$
and the 
function $\delta^*$ as defined in the context of a nondeterministic
transducer below.
\end{defi}

\begin{rem}\rm
This definition of aperiodicity for a DFA is the usual one
(see \cite{ste85}).
For a NFA,
a statement obviously equivalent to Definition \ref{defaper}
would be that $A$ is aperiodic iff applying the
subset construction to $A$ yields an aperiodic DFA.
Hence \cite{sch65} a language $L$ is star-free iff some aperiodic
(deterministic or nondeterministic) finite automaton accepts $L$.
\end{rem}

We now prepare the ground for the main result of this section,
namely that single-valued aperiodic nondeterministic finite
transducers characterize $\FO$-translations.

%
%

%
%
%

\begin{defi}
A {\em finite transducer} is given by a set $Q$ of {\em states}, an {\em
  input alphabet} $\Sigma$, an {\em output alphabet} $\Gamma$, an {\em
  initial state} $q_0$, a {\em transition relation} $\delta \subseteq Q
  \times \Sigma \times \Gamma \times Q$ and a set $F\subseteq Q$ of
  {\em final states}. For a string 
  $w=w_1\cdots w_n\in\Sigma^*$ we define the {\em set $O_M(w)$ of
  outputs of $M$ on 
  input $w$} as follows. A string $v\in\Gamma^*$ of length $n$ is in
  $O_M(w)$, if 
  there is a sequence $s_0=q_0,s_1,\ldots,s_n$ of states, such that $s_n\in F$
  and, for
  every $i$, $1\le i \le n$, we have $(s_{i-1},w_i,v_i,s_i) \in \delta$.

We say that $M$ is {\em single-valued} if, for every $w\in\Sigma^*$,
$|O_M(w)|=1$. If $M$ is single-valued it naturally defines a function
$f_M:\Sigma^* \rightarrow \Gamma^*$.

For every string $u\in\Sigma^*$ and every state $s\in Q$ we write
$\delta^*(s,u)$ for the set of states $s'$ that are reachable from $s$
on input $u$ (i.\,e., there are $s_1,\ldots,s_{|u|}=s'$ and
$v_1\cdots v_{|u|}$ such that, for 
  every $i$, $1\le i \le |u|$, we have $(s_{i-1},u_i,v_i,s_i) \in
  \delta$).

As per Definition \ref{defaper},
$M$ is {\em aperiodic} if there is an $n\in\nat$ such that for all
states $q$ and all strings $w$, $\delta^*(q,w^n)=\delta^*(q,w^{n+1})$.
\end{defi}

We will need some basic properties of FO-logic on strings.

Let $k$ be a fixed natural number and $\Sigma$ an alphabet. For every
string $u$ we write $\Phi_u^k$ for the set of 
FO-sentences of 
quantifier-depth $k$ that hold in $u$. Let $S^k$ denote the set
$\{\Phi_u^k \mid u\in\Sigma^*\}$. It is well-known that $S^k$ is
finite, for every fixed $k$ and $\Sigma$.


\begin{lemma} \label{lem:concat}
  Let $u,u',v,v'$ be strings such that $\Phi_u^k=\Phi_{u'}^k$ and
  $\Phi_v^k=\Phi_{v'}^k$. Then $\Phi_{uv}^k=\Phi_{u'v'}^k$.
\end{lemma}

\begin{proof}
  As $\Phi_u^k=\Phi_{u'}^k$ and
  $\Phi_v^k=\Phi_{v'}^k$ we know that the duplicator has a winning
  strategy in the $k$-round Ehrenfeucht game on $u$ and $u'$ and in the
  game on $v$ and $v'$. These strategies can be easily combined to
  get a strategy on $uv$ and $u'v'$. From the existence of this
  winning strategy we can, in turn, conclude that
  $\Phi_{uv}^k=\Phi_{u'v'}^k$. 
\end{proof}

\begin{thm}\label{FO<->NFA}
  A function $f\colon\Sigma^*\rightarrow \Gamma^*$ is defined by an FO
  translation if and only if it is defined by a single-valued
  aperiodic finite transducer.
\end{thm}
\begin{proof}
To simplify notation we assume that
$\Gamma=(0,1)$. The proof of the general case is a straightforward
generalization. 

(only if)
Let $f\colon\Sigma^*\rightarrow \Gamma^*$ be defined by formula
$\varphi(x)$ 
of quantifier-depth $k$ (hence, for every $w\in\Sigma^*$ and every
$i\le|w|$, the $i$-th bit of $f(w)$ is 1 iff $w\models\varphi(i)$). 
We define a single-valued aperiodic finite transducer
$M$ with input alphabet $\Sigma$, output alphabet $\Gamma$, set
$S^k\times S^k\cup\{q_0\}$ of states, initial state $q_0$ and accepting
states $\{(\Phi,\Phi_\epsilon^k)\mid \Phi\in S^k\}$. Informally, 
a state $(\Phi_1,\Phi_2)$ of $M$ represents a situation, in which $M$
``knows'' that $\Phi_1$ contains exactly those formulas (of quantifier
depth $k$)  that hold in the prefix of the
input string that was already read, and it ``guesses'' that $\Phi_2$
contains exactly those formulas that hold in the remaining part of the
string.

The transition relation $\delta$ of $M$ is defined as follows. For
every $\Phi_1,\Phi_2,\Phi'_1,\Phi'_2\in S^k$, every $\sigma\in\Sigma$
and every $\tau\in\Gamma$ we let
\[
((\Phi_1,\Phi_2),\sigma,\tau,(\Phi'_1,\Phi'_2)) \in \delta,
\]
if there exist strings $u,v\in\Sigma^*$ such that $\Phi_1=\Phi_u^k$,
$\Phi'_1=\Phi^k_{u\sigma}$, $\Phi_2=\Phi^k_{\sigma v}$,
$\Phi'_2=\Phi^k_v$ and $\tau=1 \Longleftrightarrow u\sigma v \models
\varphi(|u|+1)$. 

Analogously, for every $\Phi'_1,\Phi'_2\in S^k$, every $\sigma\in\Sigma$
and every $\tau\in\Gamma$ we define
\[
(q_0,\sigma,\tau,(\Phi'_1,\Phi'_2)) \in \delta,
\]
if there exists a string $v\in\Sigma^*$ such that
$\Phi'_1=\Phi^k_{\sigma}$,
$\Phi'_2=\Phi^k_v$ and $\tau=1 \Longleftrightarrow \sigma v \models
\varphi(1)$.

We first check that $M$ is single-valued. Let $w=w_1\cdots w_n$,
and $f(w)=v_1\cdots v_n$. We set $s_0=q_0$ and, for every $i>0$,
$s_i=(\Phi^k_{w_1\cdots w_i},\Phi^k_{w_{i+1}\cdots
  w_n})$. By using Lemma \ref{lem:concat}, it is easy to verify that $s_n\in F$ and, for every $i>0$,
we have $(s_{i-1},w_i,v_i,s_i) \in \delta$. Hence $f(w)\in O_M(w)$. 

We have to show now that no string $u=u_1\cdots u_n\not=f(w)$ is in
$O_M(w)$. Assume otherwise and let $s'_0=q_0,s'_1,\ldots,s'_n$ be
a sequence of states that outputs $u$. Let, for every $i>0$,
$s'_i=:(\Psi_i,\Theta_i)$. First, it is easy to observe that, for every
$i>0$, $\Psi_i=\Phi^k_{w_1\cdots w_i}$. As $u$ is different from $v$
there must be a $j$ such that $\Theta_j\not=\Phi^k_{w_{j+1}\cdots
  w_n}$
(Note that from the definition of $\delta$ it follows that
$(s,\sigma,1,s')\in\delta$ implies $(s,\sigma,0,s')\not\in\delta$).  
We conclude that for every $i> j$,
$\Theta_i\not=\Phi^k_{w_{i+1}\cdots 
  w_n}$: Assume, otherwise that $i>j$ is minimal, such that
$\Theta_i=\Phi^k_{w_{i+1}\cdots w_n}$. 
By definition of $\delta$ and as $(s_{i-1}',w_i,\tau,s_i')\in\delta$
it follows immediately that 
$\Theta_{i-1}=\Phi^k_{w_iw_{i+1}\cdots w_n}$, a contradiction. 
Hence, in particular, $\Theta_n\neq\Phi_{\epsilon}^k$, i.e.,
$s'_n{\not\in}F$. It follows that
$M$ is single-valued and $f_M=f$.
It remains to show that $M$ is aperiodic. First of all, it is
well-known, and can be shown by an Ehrenfeucht game argument 
\cite{ebfl95}
that, for
$n=2^k$ and every $w\in\Sigma^*$ it holds
$\Phi^k_{w^n}=\Phi^k_{w^{n+1}}$. 

Let now $\Phi_1,\Phi_2,\Phi'_1,\Phi'_2\in S^k$ and let
$u,v\in\Sigma^*$ with $\Phi_1=\Phi^k_u$ and $\Phi'_2=\Phi^k_v$. From
 Lemma \ref{lem:concat} and the definition of $\delta$ we can conclude that
$(\Phi'_1,\Phi'_2)\in \delta^*((\Phi_1,\Phi_2),x)$ if and only if
$\Phi_2=\Phi^k_{xv}$ and $\Phi'_1=\Phi_{ux}$.
Hence, again with Lemma \ref{lem:concat}, we get for every $w$ the
following.
\begin{eqnarray*}
  (\Phi'_1,\Phi'_2)\in \delta^*((\Phi_1,\Phi_2),w^n)
& \Longleftrightarrow & \text{$\Phi_2=\Phi^k_{w^nv}$ and
  $\Phi'_1=\Phi^k_{uw^n}$}\\ 
& \Longleftrightarrow & \text{$\Phi_2=\Phi^k_{w^{n+1}v}$ and
  $\Phi'_1=\Phi^k_{uw^{n+1}}$}\\ 
& \Longleftrightarrow &  (\Phi'_1,\Phi'_2)\in \delta^*((\Phi_1,\Phi_2),w^{n+1})
\end{eqnarray*}
This implies that $M$ is aperiodic.

(if) 
Let $f$ be computed by a single-valued aperiodic finite
transducer $M=(Q,\Sigma,\Gamma,q_0,\delta,F)$. It is easy to check that, for
every $s,s'\subseteq Q$, the language 
\[
L(s,s')=\{u\mid s'\in \delta^*(s,u)\}
\]
is accepted by an aperiodic finite automaton. Consequently, every
$L(s,s')$ is characterized by a FO formula $\varphi^{s,s'}$. Let
$\varphi(x)$ be the formula
\[
\bigvee_{\stackrel{s,s',s'',\sigma}{s''\in F\wedge (s,\sigma,1,s')\in\delta}}
  \varphi^{q_0,s}_<(x) \wedge P_{\sigma}(x) \wedge
  \varphi^{s',s''}_>(x). 
\]
Here, for every $s$ and $s'$, $\varphi^{s,s'}_<(x)$ is the formula that is
obtained by relativizing $\varphi^{s,s'}$ to all positions that are
smaller than $x$ and $\varphi^{s,s'}_>(x)$ is the formula that is
obtained by relativizing $\varphi^{s,s'}$ to all positions that are
greater than $x$ (see for example \cite[pp.~81f]{str94}).

Hence, for every position $x$, $\varphi(x)$ becomes true in a string
$w$ if and only
if there are states $s,s',s''$ such that
\begin{itemize}
\item $M$ can reach $s$ from the initial state
  by reading the string left to $x$,
\item $M$ can reach $s'$ from $s$ by reading the symbol at
position $x$ and output a 1, and
\item $M$ can reach the final state $s''$ from $s'$
  by reading the string to right to $x$.
\end{itemize}
As $M$ is single-valued, $\varphi(x)$ defines $f_M(w)$, for every $w$.
\end{proof}

\section{First-order with groupoidal quantifiers}

\subsection{The largest attainable class: \LOGCFL}


\begin{thm}\label{logcflbit}
There is a fixed groupoid $G$ such that
$$\QG\FO\bit = \FO\bit(\QGrp) = \LOGCFL.$$
\end{thm}

\begin{proof}
$\QG\FO\bit \subseteq \FO\bit(\QGrp)$ holds by definition for any
groupoid $G$.  To see that $\FO\bit(\QGrp) \subseteq \LOGCFL$, note
that \cite[Theorem 8.1]{baimst90} implies 
the existence of a logspace-uniform
$\ACn$-reduction, from any language in $\FO\bit(\QGrp)$, to a set of
groupoid word problems.  The unbounded fan-in AND gates in the $\ACn$
reduction can be replaced by log depth bounded fan-in sub-circuits.
Then the groupoid word problem oracle gates, of which no more than a
constant number can appear on any path from circuit inputs to circuit
output,
can be expanded into
$\SACe$ sub-circuits, since groupoid word problems are
context-free languages.
There results a logspace-uniform $\SACe$ circuit, proving membership
in \LOGCFL.

$\LOGCFL\subseteq \QG\FO\bit$ is seen by appealing to the fixed $G$
whose word problem is \LOGCFL-complete under DLOGTIME reducibility
\cite{belemc93}.
Since DLOGTIME was
shown expressible in $\FO\bit$ by \cite{baimst90}, the inclusion
follows.
\end{proof}

\subsection{Capturing \LOGCFL\ without BIT}

\begin{thm} \label{logcflnobit}
There is a fixed groupoid $G$ such that
$\LOGCFL\subseteq\QG\FO$.
\end{thm}

\begin{proof}
We first show how to express plus and times and their negations
as $\posFO(\QGrp)$ formulas (i.\,e.~formulas which have outside
of the groupoidal quantifier only a first-order
quantifier prefix and in particular
no negation).

Let us look at the predicate ``$a\cdot b=c$.'' Define
$L\eqdef\set{w\in(0,1,\#)^*}{|w|_0=|w|_1}$ and
\[\phi(a,b,c) \eqdef 
Q_L(x,y,z)\bigl[(z={\rm min})\wedge (x\leq a) \wedge (y\leq b),\ 
                (z=y={\rm max})\wedge (x\leq c)\bigr].\]
Given a word $w$ of length $n$ and assignments for $a,b,c$, the
transformation 
$[z={\rm min}\wedge x\leq a \wedge y\leq b, z=y={\rm max}\wedge x\leq c]$
yields a string of length $n^3$ over the alphabet $(0,1,\#)$ which 
contains $a\cdot b$ many $0$s, $c$ many $1$s, and $n^3-ab-c$ many
$\#$s. Thus this image is in $L$ if and only if $a\cdot b=c$.

Observe that $L$ is deterministic context-free, therefore its complement
is context-free and we conclude that
we can also express $a\cdot b\neq c$ by a $\posFO(\QGrp\FO)$ formula
(in fact even by a $\QGrp\FO$ formula).

In a similar way we can express $a+b=c$ and $a+b\neq c$ by
$\posFO(\QGrp)$ formulas.
All context-free languages involved in the definition of these
predicates can be combined into one language $L_0$, which is context-free
and co-context-free.
Now integer addition and multiplication are enough to simulate the
BIT predicate.
Indeed it can be shown that exponentiation can be defined from
addition and multiplication (see e.\,g.~\cite[p.~301]{hapu93}
and \cite[p.~192]{smo91}),
and from this it is not so hard to define the BIT predicate, as pointed
out by \cite{lin94} (cf.,~\cite{imm98}).
We conclude that there is a $\posFO(\QGrp)$ formula for the bit
predicate. The only groupoid quantifiers needed in this definition are
$Q_{L_0}$ quantifiers, and they are applied to quantifier-free formulas.

{From} Theorem~\ref{logcflbit} we know that $\LOGCFL=\QGrp\FO\bit$. 
Thus every set in $A\in\LOGCFL$ can be defined by a formula
\begin{equation}\label{f0}
Q_L\ol{x}\bigl[\Phi_1,\dots,\Phi_s\bigr],
\end{equation}
where each $\Phi_i$ is a $\FO\bit$ formula.

We will show how every such formula can be transformed into
$Q_{L'}\FO$--formula, for some fixed context--free language $L'$. 

Using the argument above we can replace each $\Phi_i$  in~(\ref{f0})
by a formula without bit,
but using the $Q_{L_0}$ quantifier.
This formula can then be transformed into the form
\begin{equation}\label{f1}
  \exists\ol{x}_1\forall\ol{x}_2\exists\ol{x}_3\cdots
        \bigvee_{i_1}\bigwedge_{i_2}\phi_{i_1,i_2},
\end{equation}
where each of the $\phi_{i_1,i_2}$ is either a positive atomic formula
or a formula of the form $Q_{L_0}\chi$, where $\chi$
is quantifier-free.

Now we combine stepwise the inner quantifiers $Q_{L_0}$ ($1\leq j\leq m$)
in formula (\ref{f1}) with the first-order connectives $\bigvee$, $\bigwedge$
and the first-order quantifiers $\exists$, $\forall$.
We give the construction for the case of an existential quantifier.
Consider the formula 
$\exists x Q_{L_1}\ol{y} \bigl[\xi_1,\dots,\xi_{k-1}\bigr]$,
where $L_1\seq A^*$ is context-free and co-context-free.
Suppose $A=(a_1,\dots,a_k)$, $\#\not\in A$.
Let $\ol{y}=(y_1,\dots,y_l)$.
This formula is equivalent to
$Q_{L_2}(x,z,y_1,\dots,y_l)\bigl[\xi_0,\xi_1',\dots,\xi_{k-1}'\bigr]$
where 
\[L_2 = \set{w\in (a_1,\dots,a_k,\#)^*}%
{w=w_1\#^+w_2\#^+\cdots\#^+w_n\#^+,\ w_i\in L_1 \text{ for some }i},
\]
$\xi_0$ is the formula $z>1$
and each $\xi_{i}'$, $1\leq i\leq k-1$, is the formula $z=1\wedge\xi_i$.
The transformation $f$ defined by $[\xi_0,\xi_1',\dots,\xi_{k-1}']$
maps a word $w$ of length $n$ to a word $f(w)$ of length $n^{l+2}$.
$f(w)$ consists of $n$ blocks $u_1,\dots, u_n$ of length $n^{l+1}$ each:
$f(w)=u_1\cdots u_n$. Here $u_m$ corresponds to the assignment $x=m$.
Each $u_m$ consists of $n$ blocks of length $n^l$, one block for each
value of $z$. These blocks are all in $\#^*$ for $z>1$, and consist of a
word over $A$ for $z=1$. This word is exactly the word to which $w$ is
mapped under the transformation $[\xi_1,\dots,\xi_{k-1}]$, when $x=m$.
Hence we see that $f(w)\in L_2$ if there is some value $m$ such that
$u_m\in L_1\#^*$. This proves the correctness of the above construction.
Certainly $L_2$ is context-free, and since the complement of $L_1$ is
context-free, we see that the complement of $L_2$ is also context-free
(the construction of appropriate PDAs is obvious).

The combinations of a $Q_{L_j}$ with a universal quantifier,
or with a first-order connective, are dealt with analogously.

We thus replaced the sub-formulas $\Phi_i$ in formula (\ref{f0}) above
and obtained a formula of the form
\begin{equation}\label{f3}
Q_L\ol{x} \bigl[\Psi_1,\dots,\Psi_s\bigr],
\end{equation}
where each $\Psi_i$ is of the form $Q_{L_i}\psi_i$, $\psi_i$ is 
quantifier-free, and $L_i$ is context-free and co-context-free.
Let $L\seq A_0^*$, where $A_0=(a_1,\dots,a_{s})$, $\#,\$\not\in A_0$.
Let $B\eqdef (a_1,\dots,a_{s},\#,\$)$.
We now define a substitution $h$ by
\begin{align*}
h(a_1) =&\ \$L_1\#B^*   \\
h(a_2) =&\ \$\overline{L_1}\#^*L_2\#B^* \\
        &\cdots\\
h(a_i) =&\ \$\overline{L_1}\#^*\cdots\#^*\overline{L_{i-1}}\#^*L_i\#B^*\\
        &\cdots\\
h(a_{s})=&\ 
        \$\overline{L_1}\#^*\overline{L_2}\#^*\cdots\#^*\overline{L_{s-1}}\#^*
\end{align*}
and let $L'\eqdef h(L)$.
Our formula replacing~(\ref{f0}) then is
\begin{equation}\label{f4}
Q_{L'}\ol{z} \bigl[\Psi_1',\dots,\Psi_{s+1}'\bigr],
\end{equation}
where we have to construct the formulas 
$\Psi_i'$ such that the following holds:
Given a word $w$, suppose the transformation given by
$[\Psi_1,\dots,\Psi_{s-1}]$ produces for a certain assignment of the
variables $\ol{x}$ the letter $a\in A$; more specifically:
suppose that $\psi_i$ produces $w_i$ (for $1\leq i\leq s-1$).
Then $[\Psi_1',\dots,\Psi_{s+1}']$ has to produce a word
$\$w_1\#^*w_2\#^*\cdots\#^*w_{s-1}\#^*$. Certainly this can be done
with quantifier-free formulas.

Thus we have shown that $\LOGCFL\seq\QGrp\FO$.
Now define $H$ to be Greibach's hardest context-free language.
Any cfl $L$ reduces to $H$ via a homomorphism
(see \cite[p.~137]{aubebo97}.
This homomorphism is $\epsilon$-free but not length-preserving.
Applying a non-unary groupoidal quantifier to simple FO-formulas
can realize this homomorphism, provided that a new
padding or neutral symbol be introduced, to act as a filler in
any word.
Thus we see that any $Q_L\FO$ formula can be
transformed into an equivalent $Q_{pad(H)}\FO$ formula.
\end{proof}

A corollary to this proof is the following remarkable result:

\begin{coro}
Greibach's hardest context-free language with a neutral symbol
is complete for $\LOGCFL$
under quantifier-free projections without BIT.
\end{coro}

A noteworthy strengthening of
Theorem \ref{logcflbit} thus follows from Theorem \ref{logcflnobit}:

\begin{coro} \label{logcflreallynobit}
$\QGrp\FO = \FO(\QGrp) = \LOGCFL$.
\end{coro}

\subsection{Unary groupoidal quantifiers}

In the previous subsection, we have shown that the situation with
non-unary groupoidal quantifiers is clearcut, since
a single such quantifier, even without the BIT predicate, captures all
of \LOGCFL.  Here we examine the case of unary quantifiers.
In this case,
the presence or absence of the BIT predicate is once again relevant.

\subsubsection{Unary groupoidal quantifiers without BIT}

\begin{thm} \label{cfl}
$\QGrp\un\FO = \CFL$.
\end{thm}

\begin{proof}
The direction from right to left follows from \cite{belemc93}:
Every context-free language reduces via a length-preserving
homomorphism to a groupoid word problem. We can even look at the
letters in a given word as groupoid elements. This reduction can be
expressed in $\FO$. 

The direction from left to right is proved by
appealing to Theorem~\ref{Nivat} and observing that the context-free
languages have the required closure properties.
\end{proof}

%

It follows immediately that nesting unary groupoidal quantifiers (in
fact, merely taking the Boolean closure of $\QGrp\un\FO$) adds
expressiveness:

\begin{coro}
$$\begin{array}{rcl}
\QGrp\un\FO = \CFL & \subsetneq & \posBC{\QGrp\un\FO}=\posBC{\CFL} \\
                & \subsetneq  & \BC{\QGrp\un\FO}=\BC{\CFL}       \\
                & \subseteq  & \FO(\QGrp\un).
  \end{array}
$$
\end{coro}

\begin{proof}
All inclusions from left to right are clear. 
The first separation follows from the fact
that CFLs are not closed under intersection.
The second separation follows from considering the non-context-free
language $Y$, consisting of all words of the form $ww$, the
complement of which is context-free.
\end{proof}

%
%
%
%

The inclusion $\CFL\subseteq\QGrp\un\FO$ in Theorem \ref{cfl}
could have be proved alternatively by
observing that the logic $\exists M\FO$ capturing CFL 
(see \cite{lascth94}) is closed under $\FO$ translations.
We note in the same vein:

\begin{thm}
${\QGrp}\SOM=\CFL$.
\end{thm}

\begin{proof}
In \cite{lascth94} it is in fact proved that
$\CFL=\exists M \SOM$. This logic is closed under monadic second-order
($\SOM$) transformations. Hence
$\CFL\seq{\QGrp}\SOM\seq\exists M \SOM\seq\CFL$. 
\end{proof}

Can we refine Theorem \ref{cfl} and find a
universal finite groupoid $G$ which captures
all the context-free languages as $\QG\un\FO$?
Intuition from the world of monoids \cite[p.~303]{baimst90}
suggests that the answer is no.
Proving that this is indeed the case is the content of Theorem
\ref{nogo} below. 
We first make a definition and state a lemma.

Let $D_t$ be the context-free one-sided Dyck language over
$2t$ symbols, i.\,e.~$D_t$ consists of the well-bracketed words
over an alphabet of $t$ distinct types of parentheses.
Recall that a PDA is a nondeterministic
automaton which reads its input from left to right and has access to a
pushdown store with a fixed pushdown alphabet.
We say that a PDA $A$ is $k$-{\em pushdown-limited},
for $k$ a positive integer,
iff
\begin{itemize}
\item the pushdown alphabet of $A$ has size $k$, and
\item $A$ pushes no more than $k$ symbols on its stack between
any two successive input head motions.
\end{itemize}

\begin{lemma} \label{dycknogo}
No $k$-pushdown-limited PDA accepts $D_t$
when $t\geq (k+1)^k+1$.
\end{lemma}

\begin{proof}
Suppose to the contrary that a $k$-pushdown-limited PDA $A$
accepts $D_t$, where $t=(k+1)^k+1$. 
$A$ has a certain fixed number, $s$, of states.
Consider $A$'s computation as it scans a length-$n$ prefix of its
input.
Since $A$ is $k$-pushdown-limited,
no more than $(k+1)^{kn}$ different stack contents,
hence no more than $s\cdot(k+1)^{kn}$ configurations, are encountered.
But $A$ must be able to distinguish between each pair of length-$n$
prefixes consisting of left parentheses alone,
because for any two such prefixes $v_1$ and $v_2$, there
is a Dyck word $v_1w$ such that $v_2w$ is not a Dyck word.
Now, it is easy to see that $t^n$, the number of
length-$n$ words over an alphabet of $t$ left parentheses,
exceeds $s\cdot(k+1)^{kn}$ when $n$ is large.
Hence $A$ cannot accept $D_t$.
\end{proof}

\begin{thm} \label{nogo}
Any finite groupoid $G$ verifies $\QG\un\FO\subsetneq \CFL$.
\end{thm}

\begin{proof}
Suppose to the contrary that $G$ is a finite groupoid such that
$\QG\un\FO = \CFL$.
Then there is a $\FO$-translation from each context-free language to
a word problem for $G$.
This means that a finite set of PDAs
(one for each word problem ${\cal W}(\cdot,G)$) can take care of
answering each ``oracle question'' resulting from such a
$\FO$-translation.
By Theorem \ref{FO<->NFA}, each $\FO$-translation is computed by a 
single-valued NFA.
Although the NFAs differ for different context-free languages
(and this holds in particular when language alphabets differ),
the NFAs do not bolster the ``pushdown-limits'' of the PDAs which
answer all oracle questions.
Hence if $k$ is a fixed integer such that all word problems ${\cal
W}(\cdot,G)$ for $G$ are accepted by a $k$-pushdown-limited PDA,
then for any positive integer $t$,
$D_t$ is accepted by a $k$-limited-pushdown PDA.
This contradicts Lemma \ref{dycknogo} when $t=(k+1)^k+1$.
\end{proof}

In the next subsection we will see that the BIT-predicate provably
adds expressive power to the logic $\QGrp\un\FO$.
Since it is known that BIT can be expressed
either by plus and times \cite{lin94} (cf.,~\cite{imm98})
or by the majority of pairs
quantifier \cite{baimst90}, the following two simple observations about
the power of $\QGrp\un\FO$ are of particular interest.

\begin{thm}\label{majdef}
The majority quantifier is definable in $\QGrp\un\FO$.
\end{thm}

\begin{proof}
Majority is a context-free language.
\end{proof}

\begin{thm}\label{addef}
Addition is definable in $\QGrp\un\FO$.
\end{thm}

\begin{proof}
Let $i,j,k$ be positions in the input word. We want to express that
$i+j=k$. We do this by using a quantifier for the context-free
language $L\eqdef\set{0^{i-1}a1^*b0^{i-1}c1^*}{i\in\N}$. Given a word 
$w\in L$, if symbol $a$ is at position $i$ and $b$ is at position $j$,
then $c$ must be at position $i+j$.
\end{proof}

\subsubsection{Unary groupoidal quantifiers with BIT}


What are $\QGrp\un\FO\bit$ and $\FO\bit(\QGrp\un)$?  It would seem
plausible that
$\QGrp\un\FO\bit\subsetneq\FO\bit(\QGrp\un)\subset\LOGCFL$, but we are
unable to prove $\QGrp\un\FO\bit\subsetneq\LOGCFL$, much less
$\FO\bit(\QGrp\un)\subsetneq\LOGCFL$. The next lemma indicates that
proving the latter would prove $\TCn\neq\LOGCFL$, settling a major
open question in complexity theory.

\begin{lemma}\label{tcn}
$\TCn\subseteq\FO\bit(\QGrp\un)$.
\end{lemma}

\begin{proof}
$\TCn$ is captured by first-order logic with bit and majority quantifiers
\cite{baimst90}. 
\end{proof}

Hence the BIT predicate is expressive and will be difficult to defeat.
The next lemma is not surprising, but it documents the
provable expressiveness of BIT.
Recall that $\CFL=\QGrp\un\FO$ (Theorem \ref{cfl}).

\begin{lemma}
$\CFL\subsetneq\QGrp\un\FO\bit$.
\end{lemma}

\begin{proof}
The language of all words whose length is a power of two is in
$\FO\bit$ hence in
the difference of the two classes.
\end{proof}

The remainder of this subsection is devoted to documenting a more
complicated setting in which the BIT predicate provably adds
expressiveness.
We want to show that $\FO(\Qun)\subset\FO\bit(\Qun)$,
i.\,e.~that even when unary groupoidal quantifiers can be nested
arbitrarily, the BIT predicate adds strength.

For this, we define, for strings $u,w$ of equal length the operations
$\overline{u}$, $u \wedge w$ and $u \vee w$ which denote the bitwise
complementation of $u$, the bitwise AND of $u$ and $w$ and the bitwise
OR of $u$ and $w$.
We say that a string $w$ is {\em $(l,m)$-bounded} if it is in $u_1^*\cdots
u_l^*$, for some strings $u_i$ with $|u_i| \le m$, for every $i$.

We are going to make use of the following Lemma.

\begin{lemma} \label{lmbounded}
  Let $u$ be an $(l,m)$-bounded 0-1-string and $w$ an
  $(l',m')$-bounded 0-1-string,  for
  some $l,m,l',m' \ge 1$, and $|u|=|w|$. Then the following hold.
  \begin{itemize}
  \item[(a)] $\overline{u}$ is $(l,m)$-bounded.
  \item[(b)] $u \wedge w$ and $u \vee w$ are $(5(l+l'),mm')$-bounded.
  \end{itemize}
\end{lemma}

\begin{proof}
(a) is trivial. We show (b) only for  $u \wedge w$, the argument for
$u\vee w$ being completely analogous.

We show the statement by induction on $l+l'$. The induction starts
with the case $l=l'=1$. 

In this case, $u=u_1^i$ and $w=w_1^j$, for some $i, j, u_1, w_1$, with
$|u_1|\leq m$ and $|w_1|\leq m'$. 

Let $u_1 \diamond
w_1$ denote the string $u_1^{|w_1|} \wedge w_1^{|u_1|}$ of length
$|u_1||w_1|\le mm'$. Further let
$d$ and $r$ be chosen such that $|u|=d|u_1||w_1|+r$
and $r<mm'$.
Then
$u \wedge w = (u_1 \diamond w_1)^dv$
for some $v$ with $|v|=r$.
Hence $u \wedge w$ is $(2,mm')$-bounded.

Now let $l+l'>2$. W.l.o.g.\ we can assume that
$u=u_1^{i_1}u_2^{i_2}u'$ and $w=w_1^jw'$ where $|u_1|,|u_2|\leq m$,
$|w_1|\leq m'$,  $u'$ is $(l-2,m)$-bounded, $w'$ is $(l'-1,m')$-bounded and
$|w_1^j| \ge |u_1^{i_1}|$.

Let $0\leq r< m'$ be such that
$|u_1^{i_1}|+r$ is a multiple of $|w_1|$.
Let $u_2^{\leftarrow r}$ be the word $u_2$ rotated $r$
positions to the left.
It should be clear that, from position $|u_1^{i_1}|+r$ in
$u\wedge w$ onwards,
the word $(u_2^{\leftarrow r} \diamond w_1)$ is repeated,
as long as the $u_2^{i_2}$ portion of $u$ and the $w_1^j$
portion of $w$ keep ``overlapping''.
We distinguish two cases.

Case 1: $|w_1^j| \le |u_1^{i_1}| + |u_2^{i_2}|$, i.e.\ the ``overlap''
with $u_2^{i_2}$ runs out within $w_1^{j}$.

\psset{unit=0.6}
\begin{center}
\begin{pspicture}(20,2)
  \psframe(2,0)(20,2)
\psline(2,1)(20,1)
\psline(8,1)(8,2)
\psline(14,1)(14,2)
\psline(11,0)(11,1)
\rput(1,1.5){$u=$}
\rput(1,0.5){$w=$}

\rput(5,1.5){$u_1^{i_1}$}
\rput(11,1.5){$u_2^{i_2}$}
\rput(17,1.5){$u'$}
\rput(6.5,0.5){$w_1^j$}
\rput(15.5,0.5){$w'$}
\end{pspicture}
\end{center}
There are $i_2',i_2'', u_3,u_4$ with $|u_3|,|u_4|<m$, such that

\begin{center}
  \begin{pspicture}(20,2)
  \psframe(2,0)(20,2)
\psline(2,1)(20,1)
\psline(8,1)(8,2)
\psline(14,1)(14,2)
\psline(11,0)(11,1)
\rput(1,1.5){$u=$}
\rput(1,0.5){$w=$}

\psline(11,1)(11,2)

\rput(9.5,1.5){$u_2^{i_2'}u_3$}
\rput(12.5,1.5){$u_4u_2^{i_2''}$}

\rput(5,1.5){$u_1^{i_1}$}
\rput(17,1.5){$u'$}
\rput(6.5,0.5){$w_1^j$}
\rput(15.5,0.5){$w'$}
\end{pspicture}

\end{center}

It is not hard to see that we can write $(u_1^{i_1}u_2^{i_2'}u_3) \wedge w_1^j$ 
as 
\[
(u_1 \diamond w_1)^{k_1} v_1v_2(u_2^{\leftarrow r} \diamond w_1)^{k_2} v_3,
\] 
for some $v_2$ of length $r$, some
$k_1,k_2$, and some $v_1,v_3$ of length at most $mm'$. As
$u_4u_2^{i_2}u'$ is $(l,m)$-bounded and $w'$ is $(l'-1,m')$-bounded it
follows by induction that $u_4u_2^{i_2}u' \wedge w'$ is
$(5(l+l'-1),mm')$-bounded. Altogether, $u \wedge w$ is
$(5(l+l'),mm')$-bounded, as required.

Case 2: $|w_1^j| \geq |u_1^{i_1}| + |u_2^{i_2}|$, i.e.\ $u_2^{i_2}$ runs
out first.

\begin{center}
\begin{pspicture}(20,2)
  \psframe(2,0)(20,2)
\psline(2,1)(20,1)
\psline(8,1)(8,2)
\psline(14,1)(14,2)
\psline(17,0)(17,1)
\rput(1,1.5){$u=$}
\rput(1,0.5){$w=$}

\rput(5,1.5){$u_1^{i_1}$}
\rput(11,1.5){$u_2^{i_2}$}
\rput(17,1.5){$u'$}
\rput(9.5,0.5){$w_1^j$}
\rput(18.5,0.5){$w'$}
\end{pspicture}
\end{center}

Hence, there are $j',j''$ and $w_2,w_3$ with $|w_2|,|w_3|<m'$ such that

\begin{center}
\begin{pspicture}(20,2)
  \psframe(2,0)(20,2)
\psline(2,1)(20,1)
\psline(8,1)(8,2)
\psline(14,1)(14,2)
\psline(17,0)(17,1)
\rput(1,1.5){$u=$}
\rput(1,0.5){$w=$}

\psline(14,0)(14,1)

\rput(5,1.5){$u_1^{i_1}$}
\rput(11,1.5){$u_2^{i_2}$}
\rput(17,1.5){$u'$}
\rput(8,0.5){$w_1^{j'}w_2$}
\rput(15.5,0.5){$w_3w_1^{j''}$}
\rput(18.5,0.5){$w'$}
\end{pspicture}
\end{center}

Now, $u_1^{i_1}u_2^{i_2} \wedge (w_1^{j'}w_2)$ can be written as 
\[
(u_1 \diamond w_1)^{k_1} v_1 v_2 (u_2^{\leftarrow r} \diamond w_1)^{k_2} v_3,
\]
where $|v_2|=r$ and $|v_1|,|v_3|<mm'$, hence this string is
$(5,mm')$-bounded. Again, by induction, it follows that the remaining
part of $u \wedge w$ is $(5(l+l'-1),mm')$-bounded, which implies the
statement of the lemma.
\end{proof}

\newcommand{\tphi}{t_{\varphi}^{\ol{y}}}

Let $\Sigma$ be a fixed alphabet, and let $\sigma$ denote the
corresponding signature.
Let $\varphi$ be a FO($+$)-$\sigma$-formula with free variables
$x$ and $\ol{y}=y_1,\ldots,y_k$. For every
string $w \in\Sigma^*$, we write $\tphi(w)$ for the 0-1 string 
$v=v_1,\ldots v_{|w|}$ with $v_i=1$ iff $\langle w,i,\ol{y}\rangle
\models \varphi$.

\begin{lemma} \label{theo:bound}
  Let $\Sigma=\{0\}$ and $\sigma_0=\{P_0\}$.
  Let $\varphi$ be a $\FO(+)$-$\sigma_0$-formula with free
  parameters $x$ and $\ol{y}=y_1,\ldots,y_k$. Then there are $l$ and $m$ such
  that for every $n$ and $y_1,\ldots,y_k$ it holds that 
$\tphi(0^n)$ is $(l,m)$-bounded.
\end{lemma}

\begin{proof}
  Let $\varphi'$ be the FO($+$)-$\emptyset$-formula which results from
  $\varphi$  by replacing every sub-formula $P_0(t)$ by true,
  introducing a new free variable, $n$, and
  restricting all quantifiers relative to $n$. I.\,e., sub-formulas
  $\exists z \theta$ are replaced by $\exists z (z<n) \wedge \theta$
  and $\forall z \theta$ is replaced by $\forall z (z<n) \rightarrow
  \theta$. Then we get
\[
\langle 0^n,x,\ol{y}\rangle \models \varphi \; \Longleftrightarrow \; 
\langle \N, n, x, \ol{y}\rangle
\models \varphi',
\]
where $\N$ denotes the natural numbers.
  Using Presburger Quantifier Elimination (see
  \cite[pp.~220ff]{boje89} or \cite[pp.~320ff]{smo87})
we can transform $\varphi'$
  into an equivalent quantifier-free formula $\psi$ which may
  additionally use the constants 0 and 1 and binary predicates $\cdot
  \equiv \cdot \; (\mbox{mod } c)$, for some constants $c$. The atomic
  formulas of $\psi$ are of one of the following forms.
  \begin{itemize}
  \item $ax + bn + a_1y_1 + \cdots a_ky_k = c$,
  \item $ax + bn + a_1y_1 + \cdots a_ky_k < c$,
  \item $ax + bn + a_1y_1 + \cdots a_ky_k > c$,
  \item $ax + bn + a_1y_1 + \cdots a_ky_k \equiv c \; (\mbox{mod } d)$,
  \end{itemize}
for some constants $a,b,c,d,a_i$.
For every fixed 
$n,y_1,\ldots,y_k$, the first formula defines, via the above
equivalence, a $(3,1)$-bounded string in $0^*1^*0^*$, the second and 
third formula define a $(2,1)$-bounded string in $1^*0^*$ and $0^*1^*$
respectively, and the last formula defines
a $(2,d)$-bounded string in $0^*(10^{d-1})^*$. As $\psi$ is fixed, by
inductively  
applying Lemma \ref{lmbounded} we get constants $l$ and $m$, such
that, for every  
$n,\ol{y}$, $\tphi(0^n)$ is
$(l,m)$-bounded. 
\end{proof}

\begin{thm}\label{bitsep2}
 $\FObit(\Qun)$ is not contained in $\FO(\Qun)$.    
\end{thm}
\begin{proof}
  We consider the language $\set{0^{n^2}}{n\in\N}$, 
which is even expressible in
  $\FObit$ and show that it is not in $\FO(\Qun)$.

In order to do so, we show that, for every unary language $L$ in
$\FO(\Qun)$, the set $\{i \;|\; 0^i \in L\}$ is semi-linear (i.\,e.~the
finite union of some arithmetic progressions).

It is enough to show that, over a one-letter
alphabet, every formula of the kind $Q_B x \varphi$ with CFL $B$ and
first-order $\varphi$ (with addition) can be replaced by
a first-order formula with addition.

Hence, let $\psi = Q_B x \varphi$, for some first-order $\phi$ (with
addition) and CFL $B$.  

Let, besides $x$, $\ol{y}=y_1,\ldots,y_k$ be the free variables of $\varphi$.

By Lemma \ref{theo:bound}, there exist $l$ and $m$ such that, for
every $n$ and $\ol{y}$, $\tphi(0^n)$ is $(l,m)$-bounded. Let
$u_1,\ldots,u_p$ be an enumeration of all 0-1 strings of length at
most $m$.  Let $L'$ denote
the (regular) language, defined by $(u_1^*\cdots u_p^*)^l$. It follows
that $\tphi(0^n)$ is in $L'$, hence it can be written as $u_{1}^{i_{11}}\cdots
u_{p}^{i_{1p}}u_{1}^{i_{21}}\cdots u_{p}^{i_{2p}}\cdots
u_{p}^{i_{lp}}$ (where, for each $j=1,\ldots,l$, all but one of the
$i_{j1},\ldots,i_{jp}$ are $0$). For a word
$w\in L'$ we write $I(w)$ for the set of tuples
$(i_{11},\ldots,i_{lp})$ with $u_{11}^{i_{11}}\cdots
u_{lp}^{i_{lp}}=w$. We show in the following that $I_B:=\bigcup_{w \in
  B \cap L'} I(w)$ is a semi-linear set.

Let $\Gamma={a_{11},\ldots,a_{1p},\ldots,a_{lp}}$ be a new
($lp$-letter) alphabet and let $h$ be the homomorphism defined by
$h(a_{ij})=u_i$. Let $\tau$ denote the Parikh mapping for strings
$a_{11}^*\cdots a_{lp}^*$. Then we have
\[
I_B = \tau(h^{-1}(B \cap L') \cap a_{11}^*\cdots a_{lp}^*),
\]
which is semi-linear by Parikh's theorem \cite[Sect.~6.9]{har78}.

Hence, $\psi$ is equivalent to a FO($+$) formula
\cite[p.~231]{har78}. By induction, we
get that every $\FO(+)(\Qun)$, hence also every $\FO(\Qun)$ formula,
over a one-letter alphabet is equivalent to a FO($+$) formula. Hence
$\set{0^{n^2}}{n\in\N}$ is not in $\FO(\Qun)$.
\end{proof}

It is interesting to see that the proof makes use of quantifier
elimination twice, first to get the bounded strings, and second to show
that $\set{0^{n^2}}{n\in\N}$ is not in FO($+$).

%

As a particular case we can now solve an open question of
\cite{baimst90}, addressing the power of different arity for majority
quantifiers.

\begin{coro}\label{coro:twoinone}
Majority of pairs can not be expressed in first-order logic with unary 
majority quantifiers.
\end{coro} 

\begin{proof}
In Theorem \ref{majdef} it was observed that the unary majority quantifier
can be simulated in $\FO(\Qun)$. On the other hand in \cite{baimst90} it
is shown that majority of pairs is sufficient to simulate the BIT
predicate. But as $\FObit(\Qun)$ is not contained in $\FO(\Qun)$ the BIT
predicate and hence the majority of pairs is not definable in $\FO(\Qun)$,
hence it cannot be simulated by unary majority quantifiers.
\end{proof}

In the same way, this time relying on Theorem~\ref{addef}, we obtain:

\begin{coro}
Multiplication is not definable in $\FO(\Qun)$.
\end{coro}

\section{Conclusion}

\begin{figure}[htb]
\begin{center}
\input{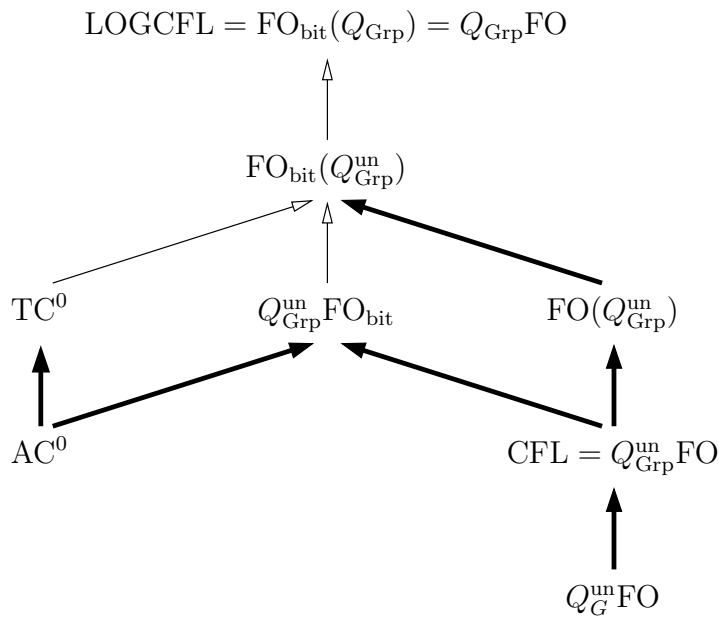}
\end{center}
\caption{The new landscape.  Here $G$ stands for any fixed groupoid,
and a thick line indicates strict inclusion.}
\label{incdiag}
\end{figure}

%
Fig.~\ref{incdiag} depicts
the first-order groupoidal-quantifier-based classes studied in this
paper.
Together with the new characterization of
$\FO$-translations by means of aperiodic finite
transducers, 
the relationships shown on Fig.~\ref{incdiag}
summarize our contribution.
%

A number of open questions are apparent from Figure \ref{incdiag}.
Clearly, it would be nice to separate the $\FO\bit$-based classes,
in particular $\FO\bit(\QGrp\un)$ from $\FO\bit(\QGrp)$,
but this is a daunting task.
A sensible approach then is to begin with $\QGrp\un\FO\bit$.
How does this compare with $\TCn$ for example?
Can we at least separate $\QGrp\un\FO\bit$ from $\LOGCFL$?
We know that $\QGrp\un\FO\bit\not\seq\FO(\QGrp\un)$; a witness
for this is the set $\set{0^{n^2}}{n\in\N}$, cf.~the proof of
Theorem~\ref{bitsep2}.

%
Other natural questions prompted by our separation results concern
extensions and refinements to Figure \ref{incdiag}.  For example, in
the world with BIT, which specific groupoids $G$ are powerful enough
to express \LOGCFL, and which are not?
In the world without BIT,
given the aperiodic transducer characterization of FO-translations,
can we prove $\REG\setminus(\REG\cap\QG\un\FO)\neq\emptyset$
as easily as Lemma \ref{dycknogo} implies
$\CFL\setminus\QG\un\FO\neq\emptyset$?
More importantly,
can we hope for an algebraic theory of groupoids to explain the detailed
structure of CFL, much in the way that an elaborate theory of monoids
is used in the extensive first-order parameterization of REG?

But perhaps the most fundamental (and hopefully tractable) question arising
from our work is not apparent from Figure \ref{incdiag}.
It concerns the Boolean closure of the context-free languages.
We have trivially used $\BC{\CFL}$ (in fact
$\posBC{\CFL}$ sufficed) to witness the separation between
$\QGrp\un\FO$ and $\FO(\QGrp\un)$.
But what is $\BC{\CFL}$ exactly,
and what techniques are available to prove that a language is not in
$\BC{\CFL}$?  It is easy to prove that any non-regular language over a
unary alphabet does not belong to $\BC{\CFL}$, and a natural infinite
hierarchy within $\posBC{\CFL}$ is known \cite{liwe73}, but the full
question seems to have fallen into the cracks.
We have several good
candidates for membership in $\FO(\QGrp\un)\setminus\BC{\CFL}$, but so
far have been unable to prove these two classes different.

Finally, ever since the regular languages in $\ACn$ and in $\ACCn$ were
characterized (the latter modulo a natural conjecture
\cite{bacostth92}), one might have wondered about a similar
characterization for the context-free languages in these classes, and
in $\NCe$.  A unified treatment of $\LOGCFL$ subclasses under the
banner of first-order logic might constitute a useful step towards
being able to answer these questions.  Since circuit-based complexity
classes are closed under Boolean operations however, a better
understanding of the interaction between the complement operation and
groupoidal quantifiers is required.  This once again seems to
highlight the importance of understanding $\BC{\CFL}$.

\medskip
{\bf Acknowledgments}.
We thank Dave Barrington, Gerhard Buntrock, Volker Diekert,
Klaus-J\"orn Lange,
Ken Regan, Heinz Schmitz, Denis Th\'erien,
Wolfgang Thomas, Klaus Wagner, and
Detlef Wotschke for useful discussions at one stage or another in the
course of this work.

\bibliography{komp}
\bibliographystyle{alpha}

\end{document}